# Studying the characteristics of scientific communities using individual-level bibliometrics: the case of Big Data research

Xiaozan Lyu[1,2] · Rodrigo Costas[2,3]

## Abstract

Unlike most bibliometric studies focusing on publications, taking Big Data research as a case study, we introduce a novel bibliometric approach to unfold the status of a given scientific community from an individual-level perspective. We study the *academic age*, *production*, and *research focus* of the community of authors active in Big Data research. Artificial Intelligence (AI) is selected as a reference area for comparative purposes. Results show that the academic realm of "Big Data" is a growing topic with an expanding community of authors, particularly of *new authors* every year. Compared to AI, Big Data attracts authors with a longer *academic age*, who can be regarded to have accumulated some publishing experience before entering the community. Despite the highly skewed distribution of productivity amongst researchers in both communities, Big Data authors have higher values of both *research focus* and *production* than those of AI. Considering the community size, overall academic age, and persistence of publishing on the topic, our results support the idea of Big Data as a research topic with attractiveness for researchers. We argue that the community-focused indicators proposed in this study could be generalized to investigate the development and dynamics of other research fields and topics.

**Keywords** Scientific community · Big Data research · Production · Research focus

**Mathematic Subject Classification** 97P70

**JEL Classification** O3

✉ Xiaozan Lyu
 lyuxiaozan@zucc.edu.cn

 Rodrigo Costas
 rcostas@cwts.leidenuniv.nl

1 Department of Administrative Management, School of Law, Zhejiang University City College, Hangzhou 310015, China

2 Center for Science and Technology Studies (CWTS), Leiden University, Kolffpad 1, P.O. Box 905, 2300 AX Leiden, The Netherlands

3 Centre for Research on Evaluation, Science and Technology (CREST), Stellenbosch University, RW Wilcocks Building, 7600 Stellenbosch, South Africa

## Introduction

"Scientific community" is common notion in the study of science, largely due to Kuhn's use of the term in his highly influential book *The Structure of Scientific Revolutions* (Jacobs, 2006). The scientific community indicates the community of people that generate scientific ideas, publish scientific journals, organize conferences, train scientists, distribute research funds, etc. (Darwin & Mendel, 2013). Kuhn (1962) assigns to the concept a prominent role in his theory, pairing it with the notion of "paradigm", where knowledge in science is produced and validated. This view draws our attention to the relevance of the characteristics of scientific communities as an angle to understand the development of scientific knowledge.

Bibliometric analyses can serve as a useful tool to enrich our understanding of the nature and structure of scientific communities (Wray, 2013). Recent years have witnessed an increase in the use of bibliometric approaches for evaluating academic performance of researchers based on publication data (Abramo & D'Angelo, 2011; Bornmann & Marx, 2014; Costas et al., 2010; Wildgaard et al., 2014), which provide us a better understanding of the scientific community from the individual level. Compared to bibliometric research revolving around publications, this kind of studies from the angle of individual authors highlights the relevant role of scholars in a scientific community as the primary elements in the process of knowledge production (Zalewska-Kurek et al., 2010), whose work will ultimately foster the progress of science. Besides, at a more macroscopic level, these author-level bibliometric analytics offer valuable insights for policy makers and administrators in academia on the performance of researchers with different characteristics (Abramo et al., 2016), as well as assist research managers in ascertaining the vitality and development potential of particular scientific fields (Rigby et al., 2008).

By applying various author-level indicators, bibliometric studies have confirmed the highly skewed productivity pattern in scientific publishing (e.g. Allison & Stewart, 1974; Price, 1963; Reskin, 1977), which was first formulated by Lotka (1926) as the famous inverse square law of productivity. This means that, there are large differences in publication output among scholars: a relatively small proportion of scholars contribute to the majority of the publications in science (Rørstad & Aksnes, 2015). Other studies have also provided evidence of the various factors related to the academic performance or scientific activity of researchers, among which *age* has been proved to be a key predictor of research output and impact as a central sociodemographic characteristic of scholars (Costas & Bordons, 2011; Levin & Stephan, 1991; Nane et al., 2017; Robitaille et al., 2008). The correlation of *age* and the extent of *diversification of studies* also indicates that compared to older (senior) researchers, the younger (junior) ones have a greater propensity to diversify their research activity (Abramo & D'Angelo, 2011; Abramo et al., 2018; Allison & Stewart, 1974; Robitaille et al., 2008).

Notwithstanding, less attention has been paid to examine the characteristics of the researcher groups engaged in the same topic from a community perspective, especially in the context of interdisciplinary research. Accordingly, in this paper, taking Big Data research as a case study, we introduce novel analytical approaches targeted to unfold the status of the publishing scholarly community active in a given field (i.e. Big Data research), similar to the "micro-community" in Kuhn's theory, in order to reflect the degree of development and stability of the topic from an individual point of view.

The dramatic increase in the number of publications related to Big Data has enabled bibliometricians to explore the current status of Big Data research from various perspectives, including the authorship and country-level collaboration patterns (Hu & Zhang, 2017b; Peng et al., 2017), the covered subject categories and journals in various databases (Gupta & Rani, 2019; Singh et al., 2015), the funding acknowledgement in major countries like the USA and China (Huang et al., 2016), as well as the focused topics and thematic trends among authors (Akoka et al., 2017; Huang et al., 2015; Kalantari et al., 2017; Singh et al., 2015). The results of these previous publications point to a growing popularity of Big Data research among scholars across various research areas.

In our study, we aim at knowing more about the scientific community of scholars carrying out research specifically around the topic of "Big Data", opening a novel perspective to investigate the configuration of disciplines and scientific topics. Particularly, we wonder whether the uptrend in the number of Big Data publications is also accompanied by an expansion of the *community of authors* using the term "big data" in their publications, or whether this is rather caused by a stable set of authors becoming more focused on this topic. This novel *community* perspective in the study of *scientific* fields would allow for more advanced arguments about their relevance, potential community interest, as well as their future sustainability. To be more specific, this work will elaborate on the following research questions:

**Q1:** How attractive is the topic of Big Data to researchers? What is the trend in the composition of the scientific community active in the topic over time?

**Q2:** Is the community of researchers around Big Data strongly focused on the topic? How many can be regarded as *specialists* with larger numbers of publications and a stronger focus on the topic?

## Data and methods

### Data and indicators

A comprehensive list of scholarly publications with a clear direct relationship with Big Data was obtained from the in-house Web of Science (WoS) database of the Centre for Science and Technology Studies (CWTS) in Leiden, by using the search terms "big data" and "bigdata" in title, abstract, and keywords in any year. A total of 9,596 publications have been identified ("Big Data set"). Although not all publications related to the research area of Big Data can be covered with our search strategy, such a narrow but precise approach is the most efficient in terms of unambiguously identifying publications that have the most unambiguous alignment with the core concept of "Big Data". The annual distribution of publications is shown in Fig. 1, suggesting an evident growth in the number of publications mentioning "big data" from 2008 onwards. Based on this analysis, we restricted the set of publications to those from 2008 onwards (a total of 9011 publications) for further analysis.

In this study, we define "*Big Data authors*" as the authors listed in papers clearly identifying the term "Big Data" in the title-abstract-keywords. As for "*Big Data Community*", we regard the overall set of all the *Big Data authors*. It is important to highlight here that with this approach we are adopting a narrow but precise approach: we are only interested in those authors that have explicitly used the term "big data" in their publications. It is of course possible that some authors may be related to Big Data research without mentioning it as such (e.g. by using other related denominations such as data science; or Big Data related techniques, e.g. machine learning, cloud computing, etc.). However, we argue that our narrower approach is the most efficient in terms of identifying those authors that have an unequivocal alignment with the core concept of Big Data, and not just with its different components and trends.

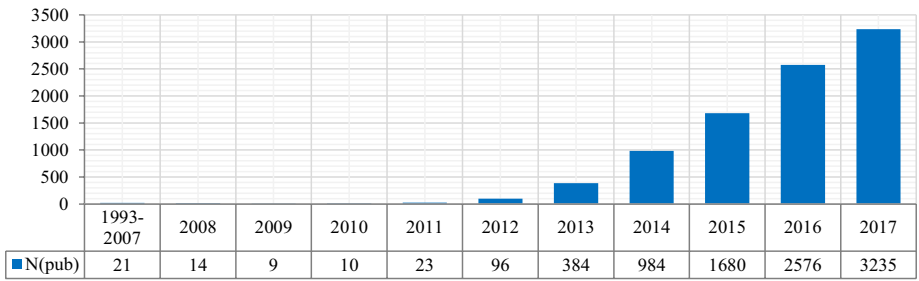

**Fig. 1** Annual number of publications related to Big Data in WoS

**Table 1** Classification and definition of authors in big data community

| Classification | Definition |
| --- | --- |
| *Old authors* | Authors who have already published papers in previous years in the Big Data set |
| *New authors* | Authors who first appeared in the Big Data set each year |
| *New-born Authors* | Authors whose first publication falls within the Big Data dataset (so they are part of the 'new authors') |
| *Stayers* | Authors who published again in Big Data in the next two years after their first publications in the topic (by definition, all stayers are 'new authors' when they publish a first Big Data paper, and then become 'old authors' in later years when they publish again) |

We use the author-name disambiguation algorithm developed by Caron and van Eck (2014) implemented in the CWTS in-house database, which links publications to one single person by using both bibliographic as well as bibliometric information. Ultimately, we work with a set of 28,922 distinct disambiguated authors. Based on the author list and years of publications, we classify the authors into several groups as shown in Table 1.

Duplicates are eliminated, which means that each author is counted only once per year, and every year there are *old authors*, *new authors,* and *new-born authors*. For instance, in the year 2012, 265 authors published a total of 96 Big Data related papers, among whom 262 (98.9%) are *new authors* without any publications on this topic before, and only 3 (1.1%) are *old authors* publishing in the previous period. Of all the 262 *new authors*, 107 (40.8%) are *new-born authors* with their first publications published in that year and the first publication being about "Big Data". Since the year 2012 is the very beginning of the boom of Big Data research (based on the number of publications indexed in WoS database), there is also a high proportion of *new authors*, especially *new-born authors*, appearing into this topic. However, it can be expected that, although there may be many *new authors* every year, only part will remain to do related research in the next years. Concerning the time window, we operationalize the *stayers* as those authors who published again in the Big Data set in the next two years after their first publication in the topic. The proportion of *stayers* captures the probability of *new authors* to become *old authors* in the next two years, which can reflect the degree of the permanent interest of authors in a topic. For the authors in 2012, about 16.2% (=43/262*100) can be considered as *stayers* who published again in Big Data set in the following two years (Table 2).

**Table 2** Number and proportion of authors in 2012 as an example

| Publish year | All authors | Old authors | New authors | New-born authors | Stayers |
|---|---|---|---|---|---|
| 2012 | 265 | 3 (1.1%) | 262 (98.9%) | 107 (40.8%) | 43 (16.2%) |

The characteristics of authors in the Big Data community are quantitatively investigated in terms of their *academic age, production,* and *research focus* through three indicators:

(1) YFP: the year of an author's first publication indexed in WoS. As the year of first publication *(YFP)* is the best single linear estimator of the ages of individual authors at the average level (Nane et al., 2017), we apply the *YFP* to characterize the *academic youth* of the authors in Big Data community (i.e. younger or senior authors). Generally, the earlier the *YFP*, the older the author and his or her seniority. The opposite would be implied to the ones with later *YFP*. Likewise, *YFP(BD)* is the first year an author engaged in Big Data community as a *new author* with a paper directly related to Big Data.

(2) $N(p)_i\_BD$: the absolute number of Big Data publications of an author in year *i*. The amount of publications that an author participates, known as the *scholarly productivity*, serves as an important measure of the rate at which they contribute units of knowledge to a certain topic (Way et al., 2017). The annual amount of these publications can be used to demonstrate the degree of activity and persistence of an author in publishing papers on Big Data.

(3) $P(f)_i\_BD$: the percentage of Big Data publications over the total output of an author in year *i*. The relative proportion of Big Data publications gives a hint of the extent to which authors are interested in the topic. In general, the higher the overall *research focus*, the more specialized the author is in the community.

The performance of individual authors in terms of *production* and *research focus* ultimately reflects the community's attention to the research field or topic, and to a certain extent, determines the potential development of the field over time. Considering the highly skewed distribution of authors in these two indicators, we simply set up 25% (i.e. dividing each distribution in top 25% and lower 75%) as the cut-off point for measuring high- and low- levels of *production* and *research focus* of authors in the community. Thus, all the authors can be categorized into four groups in a two-dimension quadrant: high/high, high/low, low/high, and low/low. An author at the high/high quadrant indicates that she or he can be regarded as *specialists* in Big Data research, with a large number of publications and high focus, while the one at the low/low quadrant is a relatively *incidental* type of author. Authors at the high/low or low/high quadrant may have a certain degree of interest in this topic. The detailed categories are as follows:

– the *specialists*: authors with both higher *production* and *research focus*, that is, both indicators fall within the top 25% intervals of the distribution of all Big Data authors;

- the *interested*: authors with lower overall *production* (lower 75%) but higher *research focus* (top 25%);
- the *casual*: authors with higher overall *production* (top 25%) but lower *research focus* (lower 75%); and
- the *incidental*: authors with both lower *production* and *research focus*, that is, both indicators fall within the lower 75% intervals.

To gain a better understanding of the relevance and community dynamics of Big Data authors, the topic of "Artificial Intelligence" (AI)[1] is selected as the reference group for further comparison. The reason for the selection of AI is that, Big Data and AI are both topics related to emerging technologies with widespread concern from various domains, and are highly esteemed in the research agendas of different governments (e.g. the *National Artificial Intelligence Research and Development Strategic Plan* of the USA). Moreover, Big Data and AI are closely related: data is the fuel that powers AI, while AI enables to make sense of massive data sets, as well as unstructured data that doesn't fit neatly into database rows and columns (Deshpande & Kumar, 2018).

For comparability, the method to identify the AI community is consistent with that for the Big Data (BD) community, *videlicet*, using the search term of "Artificial Intelligence" to retrieve all papers and the authors in WoS. Through this approach, we collected 16,835[2] disambiguated AI authors with a total of 7,427 papers during 2008–2017. Table 3 displays the generalized indicators proposed in this study, and the specific indicators based on our research objects (i.e. BD and AI communities) as well.

**Publication-level classification system**

In our study, the micro publication-level classification system of CWTS developed by Waltman and van Eck (2012) is adopted to further identify clusters for the publications and authors in our dataset at the topic level. This classification system consists of 4047 micro-clusters algorithmically generated based on citation relations among individual publications published between 2000 and 2017 in WoS. VOSviewer (van Eck & Waltman, 2010) is employed for visualization (Fig. 2). In this map, each micro-cluster contains a group of distinct publications assigned by maximizing a quality function. Therefore, each paper has only one corresponding cluster. The size of an item reflects the total number of publications in the corresponding cluster, and the distance between two items offers an approximate indication of the relatedness in terms of citation relations (van Eck & Waltman, 2017; Waltman & van Eck, 2012). Each cluster has a set of terms extracted from the titles and abstracts as its label to provide a content description (Waltman & van Eck, 2012).

By matching Big Data publications with this classification system, each paper can be allocated to a micro-cluster. The micro-clusters of authors are identical to those of their publications and are calculated in full. In this way, authors with more than one publication may have multiple clusters as the topics they engaged. For example, if an author has two publications belonging to two different clusters, this author will be counted once in each, that is, twice in total. In this sense, the micro-clusters with more authors can be considered as the core research interests of the whole community, and the micro-field with the most authors can be regarded as the main contributor to the development of Big Data research. Technically speaking, among all the 9,011 papers with a total of 28,922 authors relevant to Big Data since 2008, 7,316 (81.2%)[3] with 25,472 authors (88.1%) can be matched with micro-clusters.

---

[1] Artificial intelligence (AI) is defined "as a system's ability to correctly interpret external data, to learn from such data, and to use those leanings to achieve specific goals and tasks through flexible adaptation" (Kaplan & Haenlein, 2019).

[2] 1,127 authors engaged in both communities are counted separately.

**Table 3** Indicators used in this study

| Dimension | Abbr | Definition | BD | AI |
|---|---|---|---|---|
| Composition | $N(AU)_i$ | Number of authors in a certain field or topic in year $i$ | $N(AU)_i\_BD$ | $N(AU)_i\_AI$ |
| | $NIP(New\ AU)_i$ | Number (N) or percentage (P) of *new authors* in certain field or topic in year $i$ | $NIP(New\ AU)_i\_BD$ | $NIP(New\ AU)_i\_AI$ |
| | $NIP(New\text{-}born\ AU)_i$ | Number (N) or percentage (P) of authors published at the first time in year $i$ | $NIP(New\text{-}born\ AU)_i\_BD$ | $NIP(New\text{-}born\ AU)_i\_AI$ |
| | $NIP(Stay\ AU)_i$ | Number (N) or percentage (P) of *new authors* who published in the next two years after his/her first publication in a certain field or topic | $NIP(Stay\ AU)_i\_BD$ | $NIP(Stay\ AU)_i\_AI$ |
| | $NIP(Old\ AU)_i$ | Number (N) or percentage (P) of *old authors* in certain field or topic in year $i$ | $NIP(Old\ AU)_i\_BD$ | $NIP(Old\ AU)_i\_AI$ |
| Academic age | $YFP$ | The first year when an author published a paper in a certain field or topic indexed in WoS | $YFP(BD)$ | $YFP(AI)$ |
| Production | $N(p)_i$ | Number of publications in certain field or topic of an author in year $i$ | $N(p)_i\_BD$ | $N(p)_i\_AI$ |
| Research focus | $P(f)_i$ | Percentage of publications in certain field or topic of an author in year $i$ | $P(f)_i\_BD$ | $P(f)_i\_AI$ |

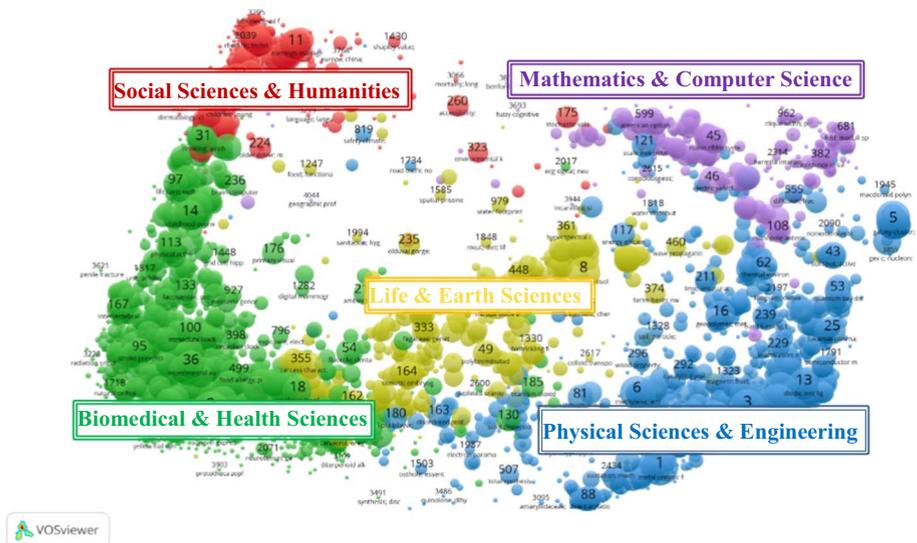

**Fig. 2** Visualization of the CWTS publication-level classification system used as a base map

# Results

In this section, perspectives of all the authors in the Big Data community (28,922 authors in 9,011 papers) and authors with research fields in the CWTS classification system (25,472 authors in 7,316 papers) are presented based on the selected indicators, including the annual number and trend of authors, as well as their characteristics in terms of the *academic age*, *production* and *research focus*, so as to provide two kinds of descriptions for the development status of this community: overall and field-oriented.

## Community of big data authors

### Size and distribution across research fields

The annual numbers of authors publishing papers in the Big Data (BD) and AI communities are displayed in Fig. 3. Despite the limited numbers of authors in the early stages, the number of Big Data authors has increased sharply since 2012, catching up with the AI community in 2014. Data for 2017 reveals that the number of authors in Big Data community is more than three times that of the AI group. This expanding trend of the author community partly explains the rapid growth of Big Data papers. Remarkably, the expansion of the number of authors in both communities is mainly due to the influx of *new authors*, being this particularly true in the AI community. Conversely, the Big Data community has a relatively larger group of *old authors* with higher degree of persistence. Moreover, the gaps between the numbers of *new* and *old authors* in both communities have gradually enlarged due to the upsurge of *new authors* over time.

---

[3] Only document types of article, letter, and review that indexed in WoS during 2000–2017 are included in this publication-level classification system (Waltman & van Eck, 2012).

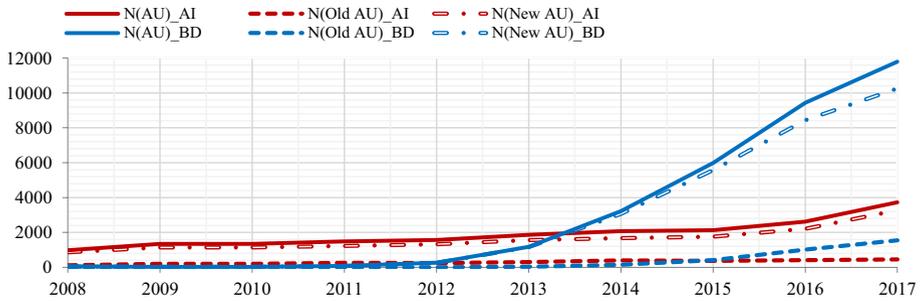

**Fig. 3** Annual number of authors in the two communities (BD and AI)

In order to better understand the distribution of Big Data authors across micro-clusters in the science landscape, we calculate the proportion of the authors in each cluster related to Big Data. The calculation at the cluster level is as follows:

$$P(AU)\_BD = \frac{N(AU)\_BD}{N(AU)}$$

Explicitly, the number of authors in a given micro cluster refers to all the distinct authors of 2008–2017 papers contained in it. The proportion distribution of Big Data authors in corresponding clusters (1,089 in all) is overlaid on the base map as shown in Fig. 4. The size of an item on the map captures the absolute number of Big Data authors in each cluster. In general, the highest proportion of authors engaged in Big Data research come from clusters in Mathematics & Computer Science (11,425, 44.9%). Following is Biomedical & Health Sciences, due to its broadest scope, constituting more than one third (8,041, 31.6%) of authors in the Big Data community. Social Sciences & Humanities ranks third in share of authors in the community (3,733, 14.7%), while Physical Sciences & Engineering and Life & Earth Sciences account for the least (9.5% and 8.8%, respectively).

Authors from Mathematics & Computer Science have the highest degree of engagement in Big Data research, in terms of the average proportion of Big Data authors in involved clusters (0.7%), far higher than those in other fields. Social Sciences & Humanities stands second with an average of 0.3% of authors publishing Big Data papers distributed in 200 clusters, followed by Life & Earth Sciences (0.2%). Authors in Biomedical & Health Sciences and Physical Sciences & Engineering have the lowest degree of engagement in Big Data community (0.1% in both cases). Specifically, clusters labeled "data science", "citizen science", "syndromic surveillance", "e-government", and "scale free networks", illustrating the largest shares of Big Data authors in each area, indicate the most active groups of authors involved in these fields. Among all the five clusters, "data science" tops the list

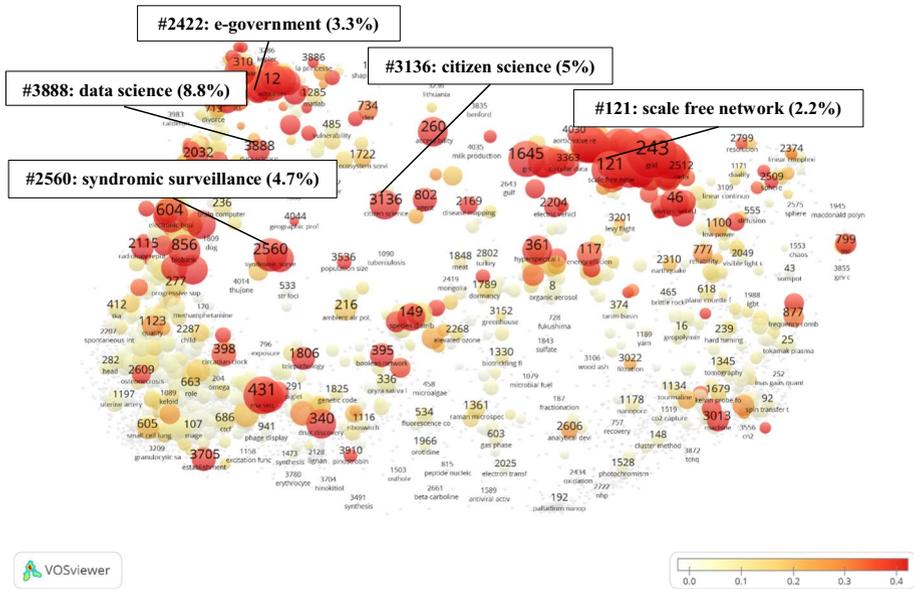

**Fig. 4** Proportion of Big Data authors [$P(AU)\_BD$] in corresponding micro-clusters. Highlighted clusters are those with the highest proportion of Big Data authors in each area

**Table 4** Statistical description of Big Data authors in research areas

| Research Areas | N(AU) | N(clusters) | Avg % of authors in clusters | Cluster with the highest proportion |
|---|---|---|---|---|
| Mathematics & Computer Science | 11,425 (44.9%) | 179 (16.4%) | 0.7 | data science (8.8%) |
| Social Sciences & Humanities | 3,733 (14.7%) | 200 (18.4%) | 0.3 | e-government (3.3%) |
| Life & Earth Sciences | 2,229 (8.8%) | 133 (12.2%) | 0.2 | citizen science (5.0%) |
| Biomedical & Health Sciences | 8,041 (31.6%) | 399 (36.6%) | 0.1 | syndromic surveillance (4.7%) |
| Physical Sciences & Engineering | 2,411 (9.5%) | 178 (16.4%) | 0.1 | scale free network (2.2%) |

Note: It is important to notice that in the publication-level classification system, publications with only a few direct citation relations may be assigned to an incorrect research area (Waltman & van Eck, 2012), such as the clusters labeled "scale free network", "citizen science", and "data science" on the map. The locations of these clusters offer insight that they are closely associated with mathematics and social science in terms of citation relations, although they are algorithmically assigned to the areas of Physical Sciences & Engineering, Life & Earth Sciences, and Mathematics & Computer Science.

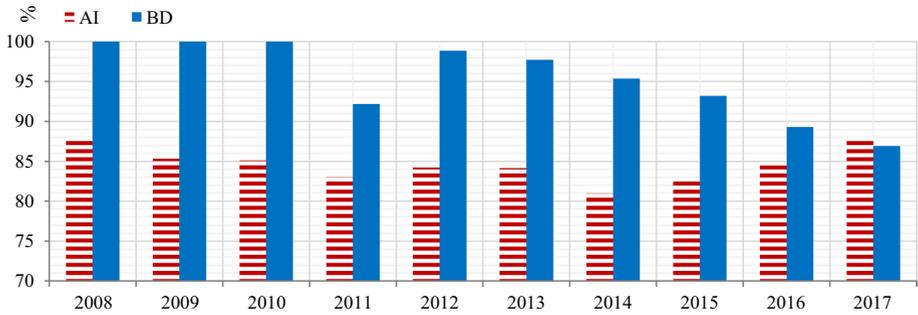

**Fig. 5** Annual percentage of *new authors* in BD and AI communities

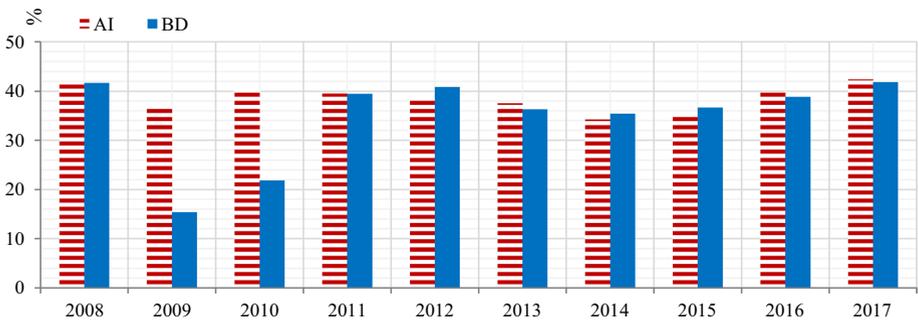

**Fig. 6** Annual percentage of *new-born authors* in BD and AI communities

with nearly 10% of its authors publishing papers directly related to Big Data (Fig. 4 and Table 4).

### New authors group

Although both topics of Big Data and AI are related to emerging technologies developing in recent years, Big Data has attracted a considerable attention as reflected by the higher proportion of *new authors* every year (about 90% on average, Fig. 5). Nevertheless, it should be noted that since the term "Big Data" has not been widely used since 2008 (Peng et al., 2017), the amount of papers directly referring to this term in the early period is relatively small, thus, nearly all the authors are *new*. Furthermore, although the numbers of *new authors* have increased year by year in both communities, the AI community has experienced an uptrend in the proportion of *new authors* since 2014, while the continuous growth in the number of *new authors* of Big Data has not led to a rise in the proportion. In fact, the opposite is true: there is a slight but constant drop in the percentage of *new authors* over time.

Figure 6 further details the annual proportions of *new-born authors* among all the *new authors*. It is revealing that except for the years 2009 and 2010, the proportions of the *new-born* in both communities are comparable: the average proportion in AI community is 38.7%, marginally higher than that in Big Data (34.8%). In addition, both communities show a slight uptrend since 2014. Such trend may hint that most (over 60%) of the authors involved in Big Data and AI research regard them as new research interests in their academic career after they have publishing experience in other topics or fields. Alternatively, we can also say that most of the Big Data and AI authors have already some level of seniority considering their previous research activity.

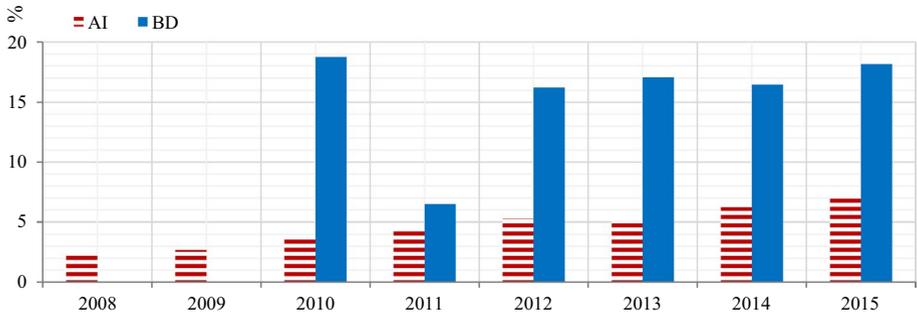

**Fig. 7** Annual proportion of *stayers* in BD and AI communities

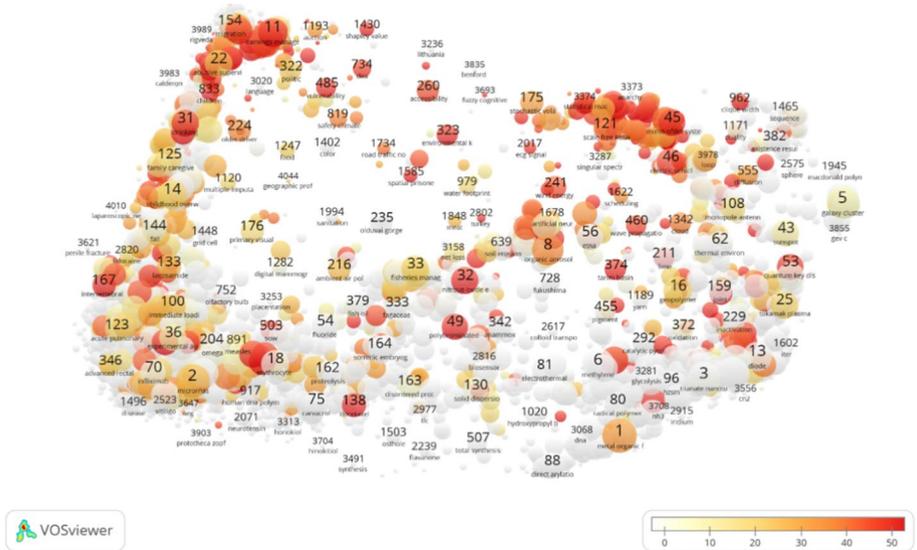

**Fig. 8** Average proportion of Big Data *stayers* across micro-clusters

As noted above, the sizes of the Big Data and AI author communities have expanded dramatically with the joint of massive new authors every year. Nevertheless, how many of these new authors will stay to do related studies in the following years? Herein, considering the time window, we calculate the percentage of *new authors* who persist in the next two years after the initial publication (i.e. they became *stayers*). Figure 7 shows that compared to AI research, the share of *stayers* is higher for Big Data research, although the share of *stayers* in AI seems to be growing steadily every year. To be more specific, since 2012, the annual proportion of *stayers* in the Big Data community has remained over 15%, much higher than that of the AI authors (about 5%).

**Table 5** Distribution of the mean *academic age (YFP)* of authors by publishing year

| PY | All authors | | New authours | | Old authors | |
|----|------|------|------|------|------|------|
|    | BD   | AI   | BD   | AI   | BD   | AI   |
| 2008 | 2000.9 | 2002.0 | 2000.9 | 2002.8 |        | 1996.2 |
| 2009 | 1999.7 | 2002.6 | 1999.7 | 2003.4 |        | 1997.5 |
| 2010 | 1999.5 | 2003.8 | 1999.5 | 2004.7 |        | 1998.3 |
| 2011 | 2002.5 | 2004.7 | 2002.7 | 2005.7 | 2000.2 | 1999.6 |
| 2012 | 2004.6 | 2005.9 | 2004.8 | 2006.6 | 1994.0 | 2001.8 |
| 2013 | 2005.3 | 2006.4 | 2005.5 | 2007.4 | 1999.3 | 2001.4 |
| 2014 | 2006.3 | 2006.8 | 2006.5 | 2007.9 | 2001.9 | 2002.5 |
| 2015 | 2007.1 | 2008.1 | 2007.4 | 2009.1 | 2002.2 | 2003.7 |
| 2016 | 2008.3 | 2009.5 | 2008.9 | 2010.3 | 2003.9 | 2004.6 |
| 2017 | 2009.6 | 2010.5 | 2010.3 | 2011.4 | 2004.9 | 2004.2 |

Next, we further analyze the overall proportion of *stayers* in their respective microclusters during 2008–2015 [*P(Stay AU)_BD*], which can offer an insight of the degree of persistence of authors in different clusters and areas (Fig. 8). The calculation at the cluster level is as follows:

$$P(StayAU)\_BD = \frac{\sum_{i=2008}^{2015} N(StayAU)i\_BD}{\sum_{i=2008}^{2015} N(NewAU)i\_BD}$$

At the research area level, the average value of *P(Stay AU)_BD*s of all the Big Data relevant clusters contained in a field is adopted to measure the general proportion of *stayers*. Overall, 11% of all the authors in the Big Data community continued to publish papers within two years after their first publications in this topic, but the distribution varies greatly. Mathematics & Computer Science has the most *stayers* (1,932, 13.0%), which consist approximately half (49.5%) of all the *stayers* in the community. More than 12% authors from Social Sciences & Humanities have continued to publish papers in the two years since becoming a *new author*, making it second in the ranking of proportion of the *stayers*. By contrast, although Biomedical & Health Sciences contains more Big Data related clusters and authors, the majority are just a sort of flash-in-the-pan in the topic, and only a small part (11.8%) continue to publish papers in the next two years since entering the community. When it comes to the clusters in Physical Sciences & Engineering and Life & Earth Science, the proportions of *stayers* are even smaller (9.68% and 5.57%, respectively).

### Academic age

The distribution of *academic age* (*YFP*) of authors by year is presented in Table 5. Comparatively, Big Data authors (all authors) have a longer *academic age* than the AI authors throughout the whole period, in other words, Big Data authors are on average elder than AI authors. This situation also applies to the other two groups – *new authors* and *old authors* in the two communities. The result corresponds to our previous finding that most Big Data authors are senior authors with already publishing experience in other topics. As mentioned above, in the two communities, the new author group accounts for an overwhelming proportion, which means that their average age greatly affects or even determines the overall academic age of the corresponding community, while the lower proportion of the old author group has a limited impact. Thus, with the narrowing age gap between the *new author* groups, the overall age gap between the two communities also has been lessening gradually.

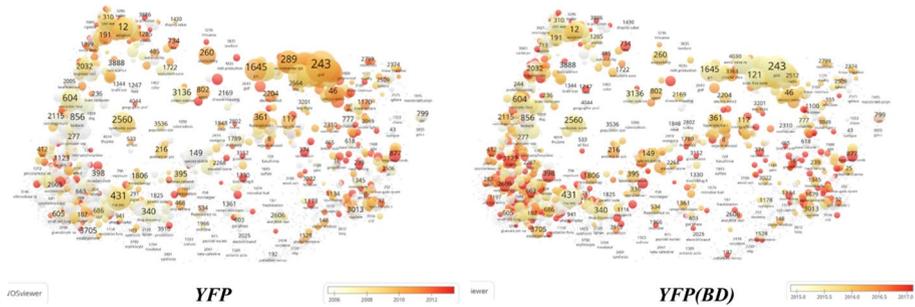

**Fig. 9** Average *YFP* and *YFP(BD)* among Big Data authors across clusters

**Table 6** Average *YFP* and *YFP(BD)* of Big Data authors in research areas

|  | Social Sciences & Humanities | Biomedical & Health Sciences | Physical Sciences & Engineering | Life & Earth Sciences | Mathematics & Computer Science |
|---|---|---|---|---|---|
| *N(AU)_BD* | 3733 | 8041 | 2411 | 2229 | 11,425 |
| *P(AU)_BD* | 14.7% | 31.6% | 9.5% | 8.8% | 44.9% |
| *YFP* | 2008 | 2006.2 | 2007.9 | 2007 | 2009.3 |
| *YFP(BD)* | 2015.8 | 2015.8 | 2016 | 2015.9 | 2015.7 |
| *YFP(BD)-YFP* | 7.8 | 9.6 | 8.1 | 8.9 | 6.4 |

The average *academic age (YFP)* and first publishing year of Big Data papers *[YFP(BD)]* of authors across clusters provide evidence that the research fields containing larger groups of Big Data authors, such as Mathematics & Computer Science and Social Sciences & Humanities, tend to have a longer overall *academic age* and began their Big Data research earlier (Fig. 9, Table 6). Put differently, authors in these clusters have a shorter time lag between the first year they published a paper (i.e. *YFP*) and the first year they had a Big Data publication (i.e. [*YFP(BD)*]). This is especially true in Mathematics & Computer Science, where authors are generally younger and are the earliest to start Big Data research, shortening the time lag within less than seven years (6.4 years). In addition, older authors usually appear in clusters in Biomedical & Health Sciences and Life & Earth Sciences. Most of these authors began to join Big Data research only in recent years with longer time gaps (more than 9 years).

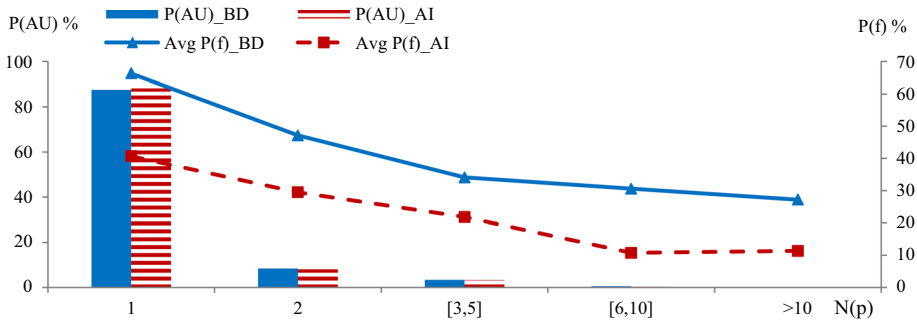

**Fig. 10** Average *research focus* of authors with different *production* levels

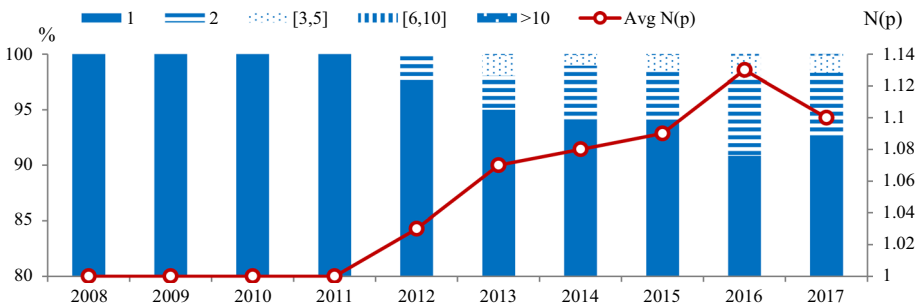

**Fig. 11** Proportions of Big Data authors (primary Y-axis) with different *production* levels (secondary Y-axis)

### Production and research focus

According to the highly skewed distributions of productivity in both communities, all the authors are divided into five groups based on their numbers of publications: only one paper, two papers, three to five papers, six to ten papers, and more than ten. The overall average *research focus [P(f)_BD]* of authors with various levels of *production [N(p)_BD]* is presented in Fig. 10. In general, the percentage distributions of authors with different levels of *production* are quite similar in the two communities: 87.5% of the author in Big Data and 88.2% authors in AI only have one paper; about 8% are authors with two publications; no more than 5% of authors published equal to or more than three papers. There are proportionally more authors in Big Data with ten or more publications (0.18%), as compared to those in AI (0.07%). Besides, although there is no strong relationship between *research focus* and *production* at the individual level, on average, the more the papers the lower the *research focus*. For those who are at the same *production* level, the Big Data authors show higher overall *focus* than the AI authors by 15 percentage point on average.

Annual proportions of authors with different *production* levels in the two communities are shown in Figs. 11, 12 respectively. The stacked bars are the proportions (primary Y-axis), and the line with dots is the corresponding average *production* of all the authors (secondary Y-axis). Each pattern corresponds to a level of *production*. Over time, since there are more authors publishing in AI earlier than 2012, their average *production* is also higher. However, Big Data authors exhibit a more rapid growth in productivity from 2012 onwards, especially those who have published two or more papers. In AI, there seems to be a decrease in average *production* from 2014 onwards, as compared to Big Data authors, who although with a slight decay in 2017 still present a high level of productivity.

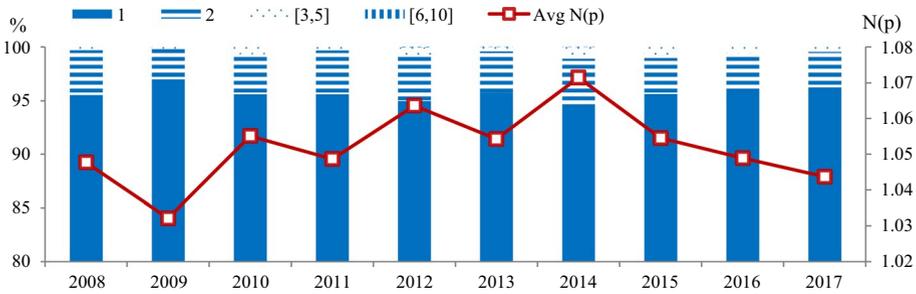

**Fig. 12** Proportions of AI authors (primary Y-axis) with different *production* levels (secondary Y-axis)

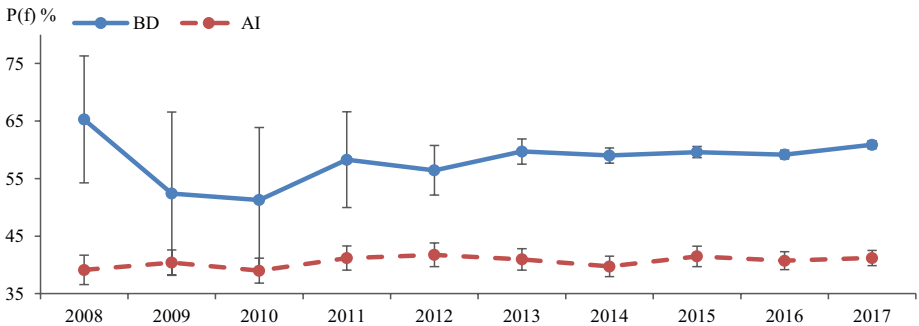

**Fig. 13** Average *research focus* of authors by year (95% CL)

With regards to the annual trend of *research focus* presented in Fig. 13, Big Data authors have an overall *research focus* of around 60%, approximately 20% more than that of the AI community (40.6%). In addition, since 2012, the annual *focus* of the Big Data community has been steadily increasing, although in the early years, due to the limited number of authors, the fluctuations are more intense. By contrast, the AI community has maintained a steadier trend, and the author's *research focus* has remained at about 40% during the period, with fluctuations of less than 2%. To sum up, Big Data authors not only have more publications, but also pay more attention to this topic, while the AI authors exhibit lower values of both overall *production* and *research focus*.

As other distributions observed in our study, the distributions of both *production* and *research focus* of Big Data authors also differ greatly across micro-clusters and research areas (Fig. 14). Computer science related clusters (e.g. clusters labeled as "grid" and "gis") are red spots in both maps, showing a relatively higher degree of *production* as well as *research focus* on Big Data. Following is the social sciences group of authors. Authors who belong to clusters within the Physical Sciences & Engineering are still quite concerned about this topic, despite their lower *production*, with an average *research focus* of more than 30%, indicating that the overall output of authors in this area is generally low. In comparison, the overlay maps provide evidence that Biomedical & Health Sciences contributes the least in Big Data research with less *production* and smaller *research focus*.

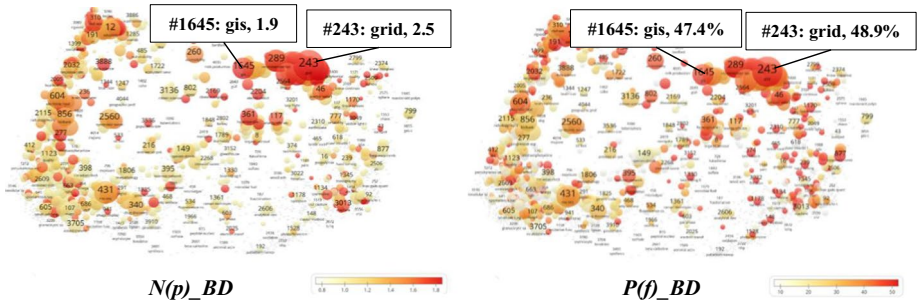

**Fig. 14** Average *production [N(p)_BD]* and *research focus [P(f)_BD]* of Big Data authors across clusters

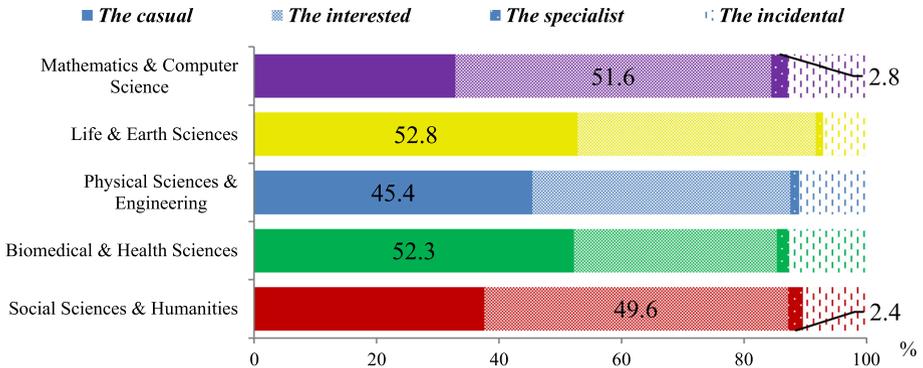

**Fig. 15** Proportion distribution of authors in different groups in research areas

In order to study the community from the prism of both *production* and *research focus*, authors have been classified depending on whether they are above or below the top 25% of the distribution in each indicator. In this way, all the Big Data authors can be divided into four non-overlapping groups as introduced in the *Indicators* part: the *specialists* (high *production*, high *research focus*), the *interested* (low *production*, high *research focus*), the *casual* (high *production*, low *research focus*), and the *incidental* (low *production*, low *research focus*).

In Big Data community, only around 2% (636) of all the authors are identified as *specialists* who published at least two papers with *research focus* of 100%. The *interested*, who also show a full *research focus* of 100%, account for the largest proportion (46.5%, 14,447) in the community, whereas they only published one paper in the whole period. The *casual* ranks second in number of authors (12,711, 42.0%), while the *incidental* group consists of about 10% (3,244) authors. Besides, though both two groups—the *casual* and *incidental*—are composed of authors with lower *research focus* at about 30%, they vary greatly in productivity: the *casual* published only one paper, while the *incidental* have an average *production* of three publications.

Referring to research areas, the proportion distribution of authors in the four groups also differs widely (Fig. 15). Almost half of authors in Mathematics & Computer Science (51.6%) and Social Sciences & Humanities (49.6%) are the *interested,* indicating their higher *research focus* though with lower *production* than the average. Besides, in comparison, Mathematics & Computer Science and Social Sciences & Humanities also have more *specialists* with better research performance (2.8% and 2.4%, respectively). Most of authors in Physical Sciences & Engineering (45.2%), Biomedical & Health Sciences (52.3%), and Life & Earth Sciences (52.8%) are the *casuals* with only one paper and lower *research focus*, around 28%.

## Discussion and conclusion

Scientific communities play an important role in the process of scientific development. In this study, we propose a bibliometric approach to investigate the composition and size of the scientific community in a given research topic. Such approach is relatively new in bibliometrics, and as suggested by Wray (2013), bibliometric approaches "have the potential to enrich our understanding of the nature and structure of scientific research communities". Therefore, this work represents a proof of concept of how the analysis of scientific communities can add an interesting color to the quantitative studies of research fields, by focusing on the composition, size and development of their publishing individuals. Given the novelty of the research topic of Big Data, in this study we developed an empirical analysis of the Big Data scientific community. We have also compared the community of Big Data authors with that of a sister area like AI in order to provide a comparative perspective.

In the context of the big data era, the year 2012 has witnessed the beginning of an exponential growth of publications related to "Big Data" from various research areas. Meanwhile, the size of Big Data scholarly community is expanding mainly due to the expansion of *new authors* every year. The percentage of *new authors* has reached over 90% on average, much higher than that in the comparative AI group, indicating a greater attractiveness to researchers. However, despite the large group of *new authors*, the *new-born* and the *stayers* both account for low proportions in the two communities, which is especially true in the AI community. This situation implies a relatively lower permanence rate of the authors in both communities, among whom the majority regard Big Data or AI as a new research interest in their academic career after they have some publishing experience in other academic topics.

With regards to the youth of authors, comparatively, Big Data authors have a longer *academic age* than the AI authors throughout the whole period. In other words, Big Data authors are by and large elder than the AI authors. However, the participation of younger *new authors* with the overwhelming proportion in Big Data community, has narrowed the overall *academic age* gap between authors in the two communities. The results provide insight that, in contrast with AI, at the early stage of development, Big Data mainly attracted *senior researchers* with already years of publishing experience, while with the maturity of this field, a large group of younger authors are also attracted to join the community.

The common phenomenon of the highly skewed distribution of productivity among authors (e.g. Allison & Stewart, 1974; Ruiz-Castillo & Costas, 2014) also exists in both communities, but is more prominent in the AI community, where a slightly larger annual proportion of one-paper authors can be observed. Correspondingly, such a large size of authors with only one paper partly explains the lower degree of both *production* and *research focus* in the AI community. By contrast, Big Data authors tend to pay more attention (i.e. *research focus*) to this topic and publish papers more frequently. Nevertheless, in the Big Data community, only a small percentage (about 2%) of authors can be identified as the *specialists* in this topic with high *production* and high *research focus*, while the overwhelming majority are the *interested* or *casual* types.

When it comes to the broader research fields, Mathematics & Computer Science stands out as the core area with the highest proportion of Big Data authors. Although younger with shorter *academic ages*, the authors in Mathematics & Computer Science joined the Big Data community the earliest in general, among whom many can be identified as senior researchers with at least two years of publishing experience in this topic. Moreover, they have larger values in terms of both *production* and *research focus*, accounting for the most significant proportion of *specialists*. Authors from Social Sciences & Humanities rank second in both average *production* and *research focus*. This area also gathers the second largest proportion of *stayers*. Physical Sciences & Engineering researchers, although gathering the lowest number of authors in Big Data research, achieved with better values in both *research focus* and *production* than the average level of the whole community, whereas those from Biomedical & Health Sciences and Life & Earth Sciences do exhibit the lowest values.

To conclude, we argue that a scientific community-based approach has the strong potential of opening up new ways for investigating the development and dynamics of research fields or topics. For instance, it would be possible to estimate the number of "big data" (or any given topic) publications in the following years based on the size of the scientific community and the average productivity of the scholars, which is more relevant than the merely analysis of publication data. Moreover, as has been proved in some previous studies (e.g., Abramo et al., 2016; Costas & Bordons, 2011; Levin & Stephan, 1991; Robitaille et al., 2008; Rørstad & Aksnes, 2015) that researchers' personal characteristics, such as age and research experience, have some relationship with their academic performance. This suggests that it could be possible to develop predictive models of the sustainability of a certain field, by considering the degrees of young and senior researchers (i.e., as measured by *YFP*) active in the field. In addition to the enrichment of the bibliometric research field, the diagnosis of the performance of scientific communities can support more advanced science policy-making, allowing policy-makers to develop more scientific community-based strategies. For example, in those strategic fields where the proportion of new or young authors is relatively low, new funding schemes for training PhDs or reinforcing collaboration between researchers from different disciplines, could be encouraged and monitored.

### Limitations and future research

It is also important to acknowledge some of the limitations of this study. Firstly, we obtain the publications on both topics (i.e. "Big Data" and "Artificial Intelligence") by using a subject search strategy with very specific terms, which means that papers related to these topics but not containing these terms are not considered. This is the same search method adopted by most bibliometric studies on "big data" research (e.g., Akoka et al., 2017; Hu & Zhang, 2017a, 2017b; Huang et al., 2016; Peng et al., 2017; Ruixian, 2013; Singh et al., 2015). However, this means that our definition of the fields and their scientific communities are strongly semantically determined. Future research should apply other delineations in order to determine potential differences in the composition of the communities based on different and more advanced delineations. However, regardless the delineation approach, it should be highlighted that the community-approach to study disciplinary domains discussed here will be essentially the same. Therefore, our results represent a proof of concept of a new approach to study the scientific community composition of disciplines from an individual point of view. Secondly, the limitation of the database chosen (WoS) as well as the document types studied (Article, Review, and Letter) also reduces the number of related papers to a large extent (Hicks, 2004). Thirdly, the author-name disambiguation algorithm and the publication-level classification adopted in this research both have their shortcomings: the clustering method used in both approaches may result in wrongly splitting the publications of an author or classifying publications in incorrect topical clusters; moreover, the cluster labels may also not always be very descriptive of the clusters (Caron & van Eck, 2014; Waltman & Van Eck, 2012).

Considering the growing presence of advanced author-name disambiguation algorithms across multiple bibliographic databases (e.g. Kim, 2018; Muelder, Faris, & Ma, 2016; Santana et al., 2017; Tekles & Bornmann, 2019), or the increasing availability of researcher identifiers such as ORCID, Scopus Researcher ID, or Clarivate's Researcher ID, it is becoming increasingly feasible the application and development of these new community perspectives to study fields dynamics. These community perspectives will allow the exploration of disciplinary dynamics from more individual-level perspectives such as gender, mobility, scientific background, funding, co-authorship networks, or team stratification (e.g. Gaule & Piacentini, 2018; Geuna & Shibayama, 2015; Larivière et al., 2009; Milojevi et al., 2019; Way et al., 2017), thus allowing for more community-based insights on the development, research potential and sustainability of research topics.

# Appendix

See Tables 7 and 8.

**Table 7** Statistical description of artificial intelligence (AI) community

| Indicators | 2008 | 2009 | 2010 | 2011 | 2012 | 2013 | 2014 | 2015 | 2016 | 2017 |
|---|---|---|---|---|---|---|---|---|---|---|
| $N(AU)_i$ | 985 | 1338 | 1343 | 1479 | 1572 | 1861 | 2071 | 2127 | 2617 | 3729 |
| $N(Old\ AU)_i$ | 120 | 196 | 200 | 251 | 248 | 295 | 394 | 371 | 406 | 460 |
| $N(New\ AU)_i$ | 865 | 1142 | 1143 | 1228 | 1324 | 1566 | 1677 | 1756 | 2211 | 3269 |
| $N(New\text{-}born\ AU)_i$ | 359 | 421 | 463 | 485 | 513 | 588 | 574 | 624 | 878 | 1386 |
| $N(Stay\ AU)_i$ | 22 | 36 | 50 | 65 | 83 | 94 | 129 | 152 | | |
| $P(Old\ AU)_i$ | 12.2 | 14.6 | 14.9 | 17.0 | 15.8 | 15.9 | 19.0 | 17.4 | 15.5 | 12.3 |
| $P(New\ AU)_i$ | 87.8 | 85.4 | 85.1 | 83.0 | 84.2 | 84.1 | 81.0 | 82.6 | 84.5 | 87.7 |
| $P(New\text{-}born\ AU)_i$ | 41.5 | 36.9 | 40.5 | 39.5 | 38.7 | 37.5 | 34.2 | 35.5 | 39.7 | 42.4 |
| $P(Stay\ AU)_i$ | 2.2 | 2.7 | 3.7 | 4.4 | 5.3 | 5.1 | 6.2 | 7.1 | | |
| $YFP(AI)_i$ | 2007.4 | 2008.3 | 2009.3 | 2010.1 | 2011.3 | 2012.3 | 2013.0 | 2014.2 | 2015.2 | 2016.3 |
| $YFP_i$ | 2002.0 | 2002.6 | 2003.8 | 2004.7 | 2005.9 | 2006.4 | 2006.8 | 2008.1 | 2009.5 | 2010.5 |
| $N(p)_i$ | 1.0 | 1.0 | 1.1 | 1.0 | 1.1 | 1.1 | 1.1 | 1.1 | 1.0 | 1.0 |
| $P(f)_i$ | 39.1 | 40.4 | 39.0 | 41.2 | 41.8 | 41.0 | 39.7 | 41.5 | 40.7 | 41.2 |

**Table 8** Statistical description of big data (BD) community

| Indicators | 2008 | 2009 | 2010 | 2011 | 2012 | 2013 | 2014 | 2015 | 2016 | 2017 |
|---|---|---|---|---|---|---|---|---|---|---|
| $N(AU)_i$ | 48 | 26 | 32 | 77 | 265 | 1178 | 3219 | 5982 | 9438 | 11,799 |
| $N(Old\ AU)_i$ | | | | 6 | 3 | 27 | 150 | 406 | 1008 | 1544 |
| $N(New\ AU)_i$ | 48 | 26 | 32 | 71 | 262 | 1151 | 3069 | 5576 | 8430 | 10,255 |
| $N(New\text{-}born\ AU)_i$ | 20 | 4 | 7 | 28 | 107 | 418 | 1087 | 2043 | 3274 | 4292 |
| $N(Stay\ AU)_i$ | | | 6 | 5 | 43 | 201 | 529 | 1086 | | |
| $P(Old\ AU)_i$ | | | | 7.8 | 1.1 | 2.3 | 4.7 | 6.8 | 10.7 | 13.1 |
| $P(New\ AU)_i$ | 100 | 100 | 100 | 92.2 | 98.9 | 97.7 | 95.3 | 93.2 | 89.3 | 86.9 |
| $P(New\text{-}born\ AU)_i$ | 41.7 | 15.4 | 21.9 | 39.4 | 40.8 | 36.3 | 35.4 | 36.6 | 38.8 | 41.9 |
| $P(Stay\ AU)_i$ | | | 18.8 | 6.5 | 16.2 | 17.1 | 16.4 | 18.2 | | |
| $YFP(BD)_i$ | 2008.0 | 2009.0 | 2010.0 | 2010.9 | 2011.9 | 2013.0 | 2013.9 | 2014.9 | 2015.8 | 2016.7 |
| $YFP_i$ | 2000.9 | 1999.7 | 1999.5 | 2002.5 | 2004.6 | 2005.3 | 2006.3 | 2007.1 | 2008.3 | 2009.6 |
| $N(p)_i$ | 1.0 | 1.0 | 1.0 | 1.0 | 1.0 | 1.1 | 1.1 | 1.1 | 1.1 | 1.1 |
| $P(f)_i$ | 65.3 | 52.4 | 51.3 | 58.3 | 56.4 | 59.7 | 59.0 | 59.6 | 59.2 | 60.9 |

**Acknowledgements** Xiaozan Lyu was supported by China Scholarship Council (CSC Student ID 201806320214), and the National Nature Science Foundation of China (NSFC, Grant Number: 71843012). Rodrigo Costas was partially supported by funding from the DST-NRF Centre of Excellence in Bibliometrics and Science, Technology and Innovation Policy (SciSTIP) (South Africa).